\documentclass[12pt]{amsart}
\usepackage{amssymb,verbatim,amscd,graphicx}

\theoremstyle{plain}
\newtheorem{thm}{Theorem}
\newtheorem{lem}[thm]{Lemma}

\newtheorem{cor}[thm]{Corollary}

\theoremstyle{definition}

\newtheorem*{rem}{Remark}

\newcommand\R{\mathbb R}
\newcommand\C{\mathbb C}

\newcommand\diag{\operatorname{diag}}

\newcommand\rank{\operatorname{rank}}
\newcommand\Flat{\operatorname{Flat}}

\begin{document}

\title[Tree Identifiability]{The identifiability of tree topology
for phylogenetic models, including covarion and mixture models}

\author[E.~S.~Allman]{Elizabeth S. Allman}
\author[J.~A.~Rhodes]{John A. Rhodes}

\date{November 7, 2005}

\address{ESA: Department of Mathematics and Statistics\\University of
Alaska Fairbanks\\Fairbanks, AK 99775} \email{e.allman@uaf.edu}

\address{JAR: Department of Mathematics and Statistics\\University of
Alaska Fairbanks\\Fairbanks, AK 99775} \email{j.rhodes@uaf.edu}

\begin{abstract}
For a model of molecular evolution to be useful for phylogenetic
inference, the topology of evolutionary trees must be identifiable.
That is, from a joint distribution the model predicts, it must be
possible to recover the tree parameter.

We establish tree identifiability for a number of phylogenetic
models, including a covarion model and a variety of mixture models
with a limited number of classes.

The proof is based on the introduction of a more general model,
allowing more states at internal nodes of the tree than at leaves,
and the study of the algebraic variety formed by the joint
distributions to which it gives rise. Tree identifiability is first
established for this general model through the use of certain
phylogenetic invariants.
\end{abstract}

\maketitle

\section{Introduction}

In phylogenetics, probabilistic models of the evolution of
biological sequences (DNA or proteins, for example) are used to
infer evolutionary history.

The parameters of such a model typically include such things as the
topology of the rooted tree depicting the temporal ordering of
speciation events, the elapsed time between these events, and the
rates at which different types of substitutions ($A\to C$, $A\to G$,
etc.) occur between events. While any of these parameters might be
of interest in a particular study, the tree topology is often the
one of greatest interest (and one on which the very definition of
the others depends).

A basic question concerning any statistical model then is whether it
is \emph{identifiable}: Given a distribution of observations that
the model predicts, is it theoretically possible to recover the
parameters of the model? Understanding what parameters are
identifiable for a model is crucial to understanding what we may
reasonably hope to infer from data.

In particular, identifiability of the tree topology is essential for
any model that is to be used in inferring evolutionary histories
from data. If a tree is not uniquely determined by an expected joint
distribution, then one has no hope of using the model to infer trees
well from data. Indeed, proofs of the statistical consistency of an
inference method such as maximum likelihood begin by establishing
identifiability of parameters.

\smallskip

Identifiability of tree topologies has been investigated for a
number of models. See, for example,  \cite{CH91, SSH94, MR97k:92011,
WadSt97, MR1664261, MR2000e:92016,Rog01} for both positive and
negative results. However, much remains to be done. As pointed out
in \cite{MEPLikelihood}, for mixture models that allow several
classes of sites in sequences to evolve at different rates,
``[n]othing has been proved in the general context yet.'' In fact,
most proofs of tree identifiability for various models have been
based on notions of phylogenetic distances and the four-point
condition of Buneman \cite{Bun}, and for general mixture models no
such distance is known. (See also \cite{EW04} for related work
illustrating non-identifiability of edge lengths for mixture
models.)

\medskip

In this paper we establish a general identifiability result
applicable to many mixture models, albeit  with a limited number of
classes. As further motivation for our work, however, we choose to
highlight the \emph{covarion model}, for which the question of
identifiability of trees has also been open.

A covarion model of character evolution describes characters with
states that are only partially observable. Such a model can be
viewed as a type of hidden Markov model on a tree, where not only
are character states at all internal nodes of the tree hidden, as in
simpler phylogenetic models, but even at the leaves full information
is not available.

For instance, in the covarion model of Tuffley and Steel
\cite{MR1604518} for the evolution of DNA sequences, each site in
the sequences is a character. As evolution proceeds over a tree,
this character is in one of the eight states $A_{\text{\emph{on}}}$,
$A_{\text{\emph{off}}}$, $G_{\text{\emph{on}}}$,
$G_{\text{\emph{off}}}$, $C_{\text{\emph{on}}}$,
$C_{\text{\emph{off}}}$, $T_{\text{\emph{on}}}$,
$T_{\text{\emph{off}}}$, where the subscript \emph{on} denotes the
site is currently free to undergo base substitution, and \emph{off}
denotes that it is currently invariable.  The off-states indicate
functional or other biological constraints temporarily preventing
substitutions. Importantly, as evolution proceeds over the tree,
sites may pass from on-states to off-states and \emph{vice versa}.
However, when we observe sequences from currently extant taxa, we
can only observe $A$, $C$, $G$, or $T$; we obtain no information as
to whether a site is currently on or off.

That constraints to nucleotide substitution might change over a tree
is a biologically plausible hypothesis that makes covarion models
attractive. A covarion model might be viewed as a type of rate
variation model, as several characters described by the same model
may be in on- and off-states for different durations, and thus
undergo different amounts of substitution. However, the `switching'
between on- and off- states, allowing a character to behave
differently in one part of a tree from another, is a crucial
distinction from standard approaches to rate variation. Of course
more elaborate covarion models, with more than the two hidden
rate-classes of the example above, are easily devised. For instance
there might be off-, slow-, and fast-states, with the opportunity
for a character to pass in and out of each as evolution proceeds
over the tree.

Though originally proposed by Fitch and Markowitz \cite{FM}, it was
not until the work of Tuffley and Steel \cite{MR1604518} that a
covarion model was mathematically formalized and the first steps
were taken in its theoretical analysis. More recently, Galtier
\cite{Galt} implemented a maximum likelihood inference package using
a covarion model, and reported improved fits to data over standard
rate-variation models. See also \cite{PMCH} for a more thorough
overview and arguments in support of the use of such models.

Although the covarion model is appealing for biological reasons, it
is less well understood theoretically. For instance, many basic
questions of identifiability of model parameters have been open.
Indeed, much of \cite{MR1604518} is focused on showing that in some
circumstances a covarion model is distinguishable from a
rate-variation model, and that some features of the tree topology
are identifiable provided one has prior knowledge of some clades.

\medskip

Motivated by the covarion model, in this paper we first prove a
result on identifiability of tree topologies for a more general
phylogenetic model. The model is introduced in Section
\ref{sec:lkgenmod}, and the result proved in Section
\ref{sec:idgen}. We also show how the result specializes to
establish identifiability of the tree topology for more specific
models of greater direct interest for applications, including the
covarion model and certain mixture models. In Sections
\ref{sec:submodels} we describe some of these models, and in Section
\ref{sec:analyticident} we apply the general result to deduce tree
identifiability under certain assumptions.

Actually, our results require some mild restrictions on model
parameters --- it is better to say that tree topology is
identifiable for \emph{generic} parameters. Informally this means if
parameters are chosen ``at random,'' then the topology can be
identified. More precisely, our use of the word ``generic'' is as in
algebraic geometry: We say a property holds for generic parameters
if it holds for all parameters off of a proper subvariety of the
parameter space. By ``subvariety'' we might mean either an algebraic
subvariety, defined by the vanishing of a set of multivariable
polynomials, or, more generally, an analytic subvariety, defined by
the vanishing of analytic functions. Since in either circumstance a
proper subvariety is a closed set of lower dimension than the
ambient space, generic parameters form an open, dense subset of the
parameter space.

\medskip

As the last paragraph hints, our approach throughout is algebraic,
and provides a good illustration of the value of an algebraic
viewpoint for statistical models. Within phylogenetics, this
approach began with the introduction of the idea of
\emph{phylogenetic invariants} by  Cavender and Felsenstein
\cite{CF87} and Lake \cite{Lake87}.  Notable contributions for
group-based models appeared in \cite{MR93m:62121} and
\cite{MR1218244}, building on an idea first introduced in
\cite{Hen89}. It has been pursued in a number of recent works
focused on phylogenetics, such as
\cite{CHHP,AR03,CKS,ARQuart,math.AG/0407033,q-bio.PE/0402015,ARSBD,ARgm,CasGarSul,CasSul,Erik},
and more broadly for biological application in the recent volume
\cite{ASCB}.

\medskip

Though our emphasis here is on theory, practical methods of
identifying tree topologies from data are also needed. For instance,
the notions of phylogenetic distance that play a key role in
theoretical identification of tree topologies for simpler models
also provide useful tools for tree inference. Whether one wishes to
base inference on a distance-based method, or merely view such
methods as fast heuristic means of finding good candidate trees to
begin a more elaborate search of tree space, the value of distances
is clear. For models where no distance formulas are known, the
explicit polynomials our results yield, whose vanishing on a joint
distribution identifies the tree topology, might play a similar
role. It will be interesting to see if these polynomials might be
exploited for practical inference, either heuristically or on a more
solid statistical basis.

\medskip

Finally, we thank Cecile An\'e for first suggesting to us that the
covarion model might be tractably studied by our methods.

\section{The $(\lambda,\kappa)$-state general Markov
model}\label{sec:lkgenmod}

In this section we introduce a phylogenetic model which allows more
states at internal nodes of the tree than at leaves. Though
motivated by the covarion model of \cite{MR1604518}, our model is
much more general. We emphasize that we introduce this model not
because we feel it precisely captures any biological phenomena, but
rather because its generality encompasses a variety of models of
more direct biological interest. It will allow us to make the key
algebraic ideas in our subsequent arguments on identifiability
clear, and obtain results which can then be applied to more
specialized models.

\medskip

Throughout, suppose $T$ is a trivalent (i.e., binary) tree. Choosing
some \emph{internal} vertex $r$ as a root, denote the resulting
rooted tree by $T^r$. Corresponding to each leaf of the tree we have
an observed random variable with state space
$[\kappa]=\{1,2,\dots,\kappa\}$, while for each internal vertex we
have a hidden (unobserved) variable with state space $[\lambda]$.
The states of observed variables might represent the bases at a site
in DNA sequences ($\kappa=4$) from extant taxa, while states of
hidden variables might represent ancestral bases together with
additional features, such as how rapidly a site currently undergoes
mutation, or even whether it is currently invariable. With this
interpretation in mind, we will always assume $\lambda\ge \kappa$.

A $\lambda$-element row vector $\boldsymbol \pi_r$ describes the
probability distribution of the states for the root variable. For
each internal edge $e$ of the tree, with $e$ directed away from the
root, a $\lambda\times \lambda$ Markov matrix $M_e$ describes
transition probabilities. For each pendant edge $e$, a
$\lambda\times \kappa$ Markov matrix $M_e$ describes transition
probabilities. Thus $M_e(i,j)$ is the conditional probability of
state $j$ at the end of $e$ given state $i$ at its start. Stochastic
assumptions ensure that all entries are non-negative, the entries of
$\boldsymbol \pi_r$ sum to 1, and the entries in any row of any of
the $M_e$ sum to 1. With no further restrictions imposed on either
the root distribution or the Markov matrices, we call this the
\emph{$(\lambda,\kappa)$-state general Markov model} on the rooted
tree $T^r$.

In the case $\lambda=\kappa$, this model is the usual general Markov
model with $\kappa$ states. We are therefore particularly interested
in cases where $\lambda>\kappa$. For instance, a generalization of
the on-off covarion model described in the introduction has
$\lambda=2\kappa$.

\medskip

For a fixed $n$-leaf rooted tree $T^r$, we may make some choice of
entries in $\boldsymbol \pi_r$ and each row of the $M_e$ to view as
independent variables, using the condition that rows sum to 1 to
determine the remaining entry. Since $T^r$ has $n$ pendant edges and
$n-3$ internal edges, the stochastic parameter space $S$ for this
model can thus be identified with a subset of $[0,1]^L$ where
$M=(\lambda-1)+n\lambda(\kappa-1)+(n-3)\lambda(\lambda-1)$.

The probabilities of observing each of the $\kappa^n$ possible
patterns (i.e., assignments of states) of leaf variables can then be
given as polynomial expressions in the parameters. That is, there is
a polynomial map, the \emph{parameterization map},
$$\phi_{T^r}:S\to [0,1]^{\kappa^n},$$
which gives the joint distribution of observed states at the leaves
of $T^r$ as a function of the parameters. We extend this to a
polynomial map $\C^L\to \C^{\kappa^n}$ which we also denote by
$\phi_{T^r}$, and refer to $\C^L$ as the \emph{complex parameter
space.}

The \emph{phylogenetic variety} for the $(\lambda,\kappa)$-state
model on $T^r$ is the algebraic variety defined as
$$V_{T^r,\lambda,\kappa}=\overline{\phi_{T^r}(\C^L)},$$ where the
bar denotes the (Zariski and standard) topological closure in
$\C^{\kappa^n}$.

\begin{lem} Let $T$ be an $n$-leaf trivalent tree, and $r_1, r_2$
any two internal nodes. Then $V_{T^{r_1},\lambda,\kappa}=V_{T^{r_2},\lambda,\kappa}$,
so this variety may be denoted by $V_{T,\lambda,\kappa}$.
\end{lem}
\begin{proof}
This is proved similarly to the corresponding result for the general
Markov model, as in \cite{SSH94} or\cite{AR03}.
\end{proof}

\medskip

\section{Algebraic and analytic submodels of the
$(\lambda,\kappa)$-state general Markov model}\label{sec:submodels}

To further motivate our introduction of the $(\lambda,\kappa)$-state
general Markov model, we note that many models of molecular
evolution can be viewed as submodels of it, in that they simply
place more restrictive assumptions on the allowable parameter
values. In this section we first indicate some of these submodels of
interest.

We also introduce the idea of an \emph{analytic
$(\lambda,\kappa)$-state model}, which will allow us not only to
deal with parameterization maps in which joint distributions are
expressed by polynomial formulas in the parameters, but a wider
class that encompasses the `rate matrix' models that are so commonly
used in applications.

\medskip

Some specific examples of models that can be viewed as submodels of
the general $(\lambda,\kappa)$-state model include the following:

\begin{enumerate}
\item GM: As already stated, the $\kappa$-state general Markov model
results from $\lambda=\kappa$.
\item GM+I: This model allows two classes of sites in sequences; one
class mutates according to the general Markov model, and another is
held invariable. A parameter $f$ denotes the proportion of sites in
the first class, with $1-f$ in the second. If the root distribution
vectors for the two classes are $\boldsymbol \pi_1$ and $\boldsymbol
\pi_2$, let $\boldsymbol \pi_r=(f\boldsymbol \pi_1,(1-f)\boldsymbol
\pi_2)$. For an internal edge $e$, if $N_e$ is the $\kappa\times
\kappa$ matrix describing transition probabilities for the first
class along  that edge, let
$$M_e=\begin{pmatrix} N_e & 0\\0&I\end{pmatrix},$$
a $2\kappa\times 2\kappa$ matrix. For pendent edges $e$, let
$$M_e= \begin{pmatrix} N_e\\I\end{pmatrix},$$
a $2\kappa\times \kappa$ matrix. Thus the model results from simply
restricting parameters in the general $(2\kappa,\kappa)$-state model
so that all Markov matrices have a particular form.

The restricted parameter space can be identified with $\C^M$, where
$M=(2\kappa-1)+(2n-3)\kappa(\kappa-1)$. Note the parameterization
map giving joint distributions as a function of parameters for this
model is still a polynomial one, given by restriction of the map for
the general $(2\kappa,\kappa)$-state model to a smaller domain.

\item GM+GM+$\cdots$ +GM: We consider $m$ classes of sites,
each evolving independently according to a different GM model. To
view this as a submodel of our more general model, for each internal
edge of the tree we create an $m\kappa\times m\kappa$ block-diagonal
Markov matrix, with each of the $m$ $\kappa\times\kappa$ blocks
giving transition probabilities for a particular class. On pendant
edges we `stack' the blocks, giving $m\kappa\times\kappa$ matrices.
The root distribution is similarly obtained by concatenating the
root distributions for each class, weighted by additional parameters
describing the relative frequencies of the classes. Thus we are
dealing with a restriction of the $(m\kappa,\kappa)$-state model,
with parameter space identified with $\C^M$ where
$M=(m\kappa-1)+(2n-3)m\kappa(\kappa-1)$. Again, the parameterization
map for this model is polynomial.

\item Other algebraic models: In the previous examples, we
can replace an occurrence of GM by a submodel, such as the
Jukes-Cantor, Kimura 2-parameter, Kimura 3-parameter, or Strand
Symmetric, defined by further restriction of parameters. Allowing
arbitrary matrices of these types on each edge (so that we do
\emph{not} assume a common rate matrix), we again have a polynomial
parameterization map with domain $\C^M$ for some $M$.

\end{enumerate}

The previous examples all lie fully within an algebraic framework,
but in fact many of the models used for inference in current
applications are not of this sort. \emph{Rate matrix models} assume
more commonality to the substitution process on the various edges of
the tree. Typically, one fixes a rate matrix $Q$ with non-negative
off-diagonal entries and rows summing to 0. Then each edge of the
tree is assigned a scalar edge length $t_e$, and the Markov matrix
$M_e=\exp(Qt_e)$ gives transition probabilities for that edge.

\begin{enumerate}

\item[(5)] GTR: A submodel of the general $(\kappa,\kappa)$-state model,
the general time-reversible model assumes a root distribution
$\boldsymbol \pi_r$ and rate matrix $Q$ such that $\boldsymbol
\pi_rQ=\boldsymbol 0$ and $\diag(\boldsymbol\pi_r)Q$ is symmetric.
Pairs $\boldsymbol\pi_r,Q$ with these properties can be
parameterized by $(\kappa-1) + (\kappa)(\kappa-1)/2$ scalars. Since
we may normalize so one edge length is 1, the parameter space for
the full model is of dimension
$M=(\kappa-1)+\kappa(\kappa-1)/2+(2n-4)$. However, the
parameterization map giving joint distributions is not polynomial,
as it involves a composition of matrix exponentials with the general
Markov parameterization. Nonetheless, it is an analytic map.

\item[(6)] GTR+\text{rate-classes}: Let $\boldsymbol \pi_r,Q$ be as in the GTR model.
Assuming $m$ different classes of sites, we assign each a relative
frequency $f_i$ and a scalar rate parameter $\lambda_i$, with
$\lambda_1=1$. Then the $i$th class undergoes substitutions on an
edge $e$ according to $N_e=\exp(Q\lambda_it_e)$. For internal edges
of the tree we embed the $N_e$ as blocks in a larger block-diagonal
matrix $M_e$, while for pendant edges we stack them, obtaining an
expression of this model as a submodel of the general
$(m\kappa,\kappa)$-state model. The parameter space is of dimension
$M=(\kappa-1)+\kappa(\kappa-1)/2+2(m-1)+(2n-4)$. Again we have an
analytic, but not polynomial, parameterization map.

Note that our formulation requires a finite number of rate classes.
While current literature often refers to a continuous distribution
of rates (usually with a $\Gamma$ distribution), in practice
inference is always done with a discretization of the distribution,
producing finitely many rate classes as here.

\item [(7)] GTR+I+\text{rate-classes}: This model is simply the last, with the
further assumption $\lambda_2=0$.

\item[(8)] Covarion:
As formulated in \cite{MR1604518}, the Tuffley-Steel covarion model
hypothesizes a common $2\kappa \times 2\kappa$ rate matrix $Q$ of a
particular form. Let $\boldsymbol \pi, R$ be a root distribution and
rate matrix for the $\kappa$-state GTR model. Let $s_1, s_2>0$, and
set $\sigma_1=\frac {s_2}{s_1+s_2}$, $\sigma_2=\frac
{s_1}{s_1+s_2}$. Then
$$Q=\begin{pmatrix} R-s_1I&s_1I\\s_2I&-s_2I\end{pmatrix}$$
is the rate matrix for a $2\kappa$ state time-reversible process,
stationary on the root distribution vector $\boldsymbol
\pi_r=(\sigma_1\boldsymbol \pi, \sigma_2\boldsymbol \pi ).$ In the
language of the introduction, the first $\kappa$ states represent
bases that are on, and the last $\kappa$ ones that are off.

The covarion model then associates to each internal edge $e$ of the
tree the Markov matrix $M_e=\exp(Qt_e)$, and to a pendant edge $e$
the matrix $M_e=\exp(Qt_e)(I_{\kappa\times \kappa}\thickspace
I_{\kappa\times\kappa})^T$. This model is therefore a submodel of
the $(2\kappa,\kappa)$-state model.

Since the parameters for the covarion model can be viewed as
$(\boldsymbol \pi,R,s_1,s_2,\{t_e\})$, the parameter space is of
dimension $M=(\kappa-1)+\kappa(\kappa-1)/2+2+(2n-4)$. As with all
rate matrix models, the parameterization map is analytic, though not
polynomial.

\item[(9)] The model referred to as the SSRV in \cite{Galt}
generalizes the covarion model to $m$ rate classes, sharing the same
rate matrix $R$, with switching allowed between the classes. It can
similarly be seen to be a submodel of the $(m\kappa,\kappa)$-state
model, with an analytic parameterization map.

\end{enumerate}

Of course many more variations are possible. The reader familiar
with other basic models will have no trouble modifying the examples
above, adding rate classes if desired, or even mixing several
different models as separate classes.

\medskip

To formalize a notion of submodel of the $(\lambda,\kappa)$-state
model that encompasses the above and other examples, we introduce
some new terminology.

By a \emph{submodel} of the $(\lambda,\kappa)$-state general Markov
model on a tree $T^r$ we mean a restriction of parameters to a
subset of the full parameter space $\C^L$. Suppose the set of
$(\lambda,\kappa)$-state general Markov model parameters $s\in \C^L$
allowed in the submodel is $\psi(U)$, the image under some analytic
map $\psi:U\to\C^L$ of an open set $U\subseteq \R^M$. Then we say
the submodel is an \emph{analytic $(\lambda,\kappa)$-state model}
with \emph{parameter space} $U$ and \emph{Markov map} $\psi$. The
\emph{parametrization map} for the analytic model is then
$\phi_{T^r}\circ \psi$, where $\phi_{T^r}$ is the parameterization
map for the general $(\lambda,\kappa)$-state model:
$$U\xrightarrow{\psi} \C^L\xrightarrow{\phi_T^r} \C^{\kappa^n}.$$

Thus, for instance, the covarion model is a analytic
$(2\kappa,\kappa)$-state model, as is the GM+I model. Note that
algebraic submodels, with polynomial Markov maps, are included among
the analytic ones. Analyticity of the Markov map will be important
for our arguments in Section \ref{sec:analyticident}.

\smallskip

While all of the enumerated models above are analytic
$(\lambda,\kappa)$-state models, they in fact have additional
structure in common. First note that in each $\lambda=m\kappa$ for
some $m$. Moreover, the set of states $[\lambda]$ at each internal
node is naturally identified with the set $[\kappa]\times [m]$.
Under this identification, if we refer to the observable states in
$[\kappa]$ as `bases', then a state $(i,j)$ represents `base $i$'
and `class $j$'. Here `class' might refer to `rate class', or some
other characteristic, such as the on/off feature in the covarion
model. Finally, in all of these models the Markov matrices on
pendant edges of the tree have a form $$\widetilde M=M (I\thickspace
I\thickspace \dots \thickspace I)^T,$$ where $M$ is an
$m\kappa\times m\kappa$ Markov matrix of the sort allowed on
internal edges. Essentially this means the model hides all class
information, so only bases are observable at leaves. We refer to
such an analytic $(m\kappa,\kappa)$-state model as an \emph{analytic
$\kappa$-base, $m$-class model}.

\section{Identifiability of the tree topology for the general
$(\lambda,\kappa)$-state model}\label{sec:idgen}

Returning to consideration of the $(\lambda,\kappa)$-state general
Markov model, in this section we establish our main technical result
on generic identifiability of tree topologies.

\smallskip

We first consider identifiability of the tree topology from a joint
distribution of states at the leaves  in the case of a 4-leaf tree.
Let the three possible trivalent trees with leaves $a,b,c,d$ be
denoted by
$$T_1=T_{ab|cd},\ T_2=T_{ac|bd},\ T_3=T_{ad|bc},$$
where the subscript $uv|wx$ denotes leaves $u,v$ are adjacent to a
common internal node, as are leaves $w,x$.

Focusing on $T_1$, denote the internal nodes by $r,f$, so that the
root $r$ is adjacent to the leaves $a$ and $b$. For $s\in \C^L$, the
complex parameter space, let the corresponding vector and matrix
parameters (with rows summing to 1) be $\boldsymbol \pi_r\in
\C^\lambda$, $M_{rf}\in M_{\lambda\times\lambda}(\C),$ $M_{ra},
M_{rb}, M_{fc}, M_{fd}\in M_{\lambda\times\kappa}(\C).$

Then $P=\phi_{T_1}(s)$ can be expressed as a
$\kappa\times\kappa\times\kappa\times\kappa$ tensor whose entries
$P(i,j,k,l)$ give the `probability' of observing states $i,j,k,l$ at
leaves $a,b,c,d$, respectively, and are given by the following
formula:

Let $A=\diag(\boldsymbol \pi_r) M_{rf}$, a $\lambda\times\lambda$
matrix. Then define a
$\lambda\times\lambda\times\lambda\times\lambda$ tensor $Q$ by
$$Q(i,j,k,l)=\begin{cases} A(i,k) &\text{ if $i=j$ and $k=l$,}\\
0&\text{ otherwise. }\end{cases}$$ Finally, let each of the $M_{ra},
M_{rb}, M_{fc}, M_{fd}$ act in the consecutive indices of $Q$ to
yield $P$, i.e.,
\begin{multline}
P(i,j,k,l)=\\
\sum_{i',j',k',l'=1}^\lambda
Q(i',j',k',l')M_{ra}(i',i)M_{rb}(j',j)M_{fc}(k',k)M_{fd}(l',l).
\label{eq:Ptens}
\end{multline}

\begin{figure}[h]
\begin{center}
\includegraphics[height=1.2in,width=4.5in]{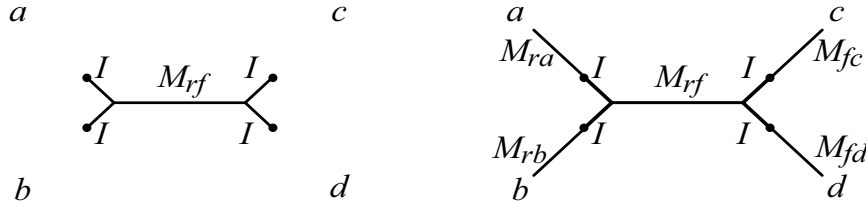}
\end{center}
\caption{Models underlying the tensors $Q$ and
$P$.}\label{fig:QandP}
\end{figure}

To motivate Equation (\ref{eq:Ptens}), and our subsequent arguments,
one should think of the matrix $A$ as expressing the joint
distribution of states at the vertices $r$ and $f$. The tensor $Q$
then represents the joint distribution for a $(\lambda,
\lambda)$-state model on a quartet tree where no state changes occur
along the pendant edges (i.e., the Markov matrices on these edges
are $I$), as illustrated at the left in Figure \ref{fig:QandP}. The
model producing $P$ `extends' these terminal edges, placing the
$\lambda\times\kappa$ Markov matrices on the extensions, as shown on
the right.

Importantly, Equation (\ref{eq:Ptens}) can be re-expressed several
ways. The first of these naturally reflects the topology of $T_1$.
Let $\Flat_{ab|cd}(Q)$ be the $\lambda^2\times\lambda^2$ matrix with
entries $$\Flat_{ab|cd}(Q)((i,j),(k,l))=Q(i,j,k,l),$$ where each
index runs through $[\lambda]^2$. Similarly let $\Flat_{ab|cd}(P)$
be the $\kappa^2\times\kappa^2$ matrix, with indices in
$[\kappa]^2$, given by
$$\Flat_{ab|cd}(P)((i,j),(k,l))=P(i,j,k,l).$$
Let $N_{ab}=M_{ra}\otimes M_{rb}$ and $N_{cd}=M_{fc}\otimes M_{fd}$
where `$\otimes$' denotes the Kronecker product of matrices. Thus
$N_{ab}$ and $N_{cd}$ are $\lambda^2\times \kappa^2$ matrices where,
for instance $N_{ab}((i,j),(k,l))=M_{ra}(i,k)M_{rb}(j,l).$ Then we
have
\begin{equation}
\Flat_{ab|cd}(P)= N_{ab}^T \Flat_{ab|cd}(Q)N_{cd}. \label{eq:FP1}
\end{equation}

Alternatively, we have other expressions involving flattenings which
are less natural with respect to the topology of $T_1$. Let
$\Flat_{ac|bd}(Q)$ be the $\lambda^2\times\lambda^2$ matrix with
entries
$$\Flat_{ac|bd}(Q)((i,j),(k,l))=Q(i,k,j,l).$$ Similarly let
$\Flat_{ac|bd}(P)$ be the $\kappa^2\times\kappa^2$ matrix with
entries
$$\Flat_{ac|bd}(P)((i,j),(k,l))=P(i,k,j,l).$$
Let $N_{ac}=M_{ra}\otimes M_{fc}$ and $N_{bd}=M_{rb}\otimes M_{fd}$.
Then we have
\begin{equation}
\Flat_{ac|bd}(P)= N_{ac}^T \Flat_{ac|bd}(Q)N_{bd}.\label{eq:FP2}
\end{equation}
A third such expression, obtained in a similar way, is
\begin{equation}
\Flat_{ad|bc}(P)= N_{ad}^T \Flat_{ad|bc}(Q)N_{bc}.\label{eq:FP3}
\end{equation}

\smallskip

The key observation underlying our proof of the identifiability of
the tree topology for the ($\lambda,\kappa$)-state model (Theorem
\ref{thm:4treeid} and Corollary \ref{cor:treeid} below) is that, for
generic parameter choices, the ranks of the matrices
$\Flat_{ab|cd}(Q)$, $\Flat_{ac|bd}(Q)$, $\Flat_{ad|bc}(Q)$, are
rather different, and this affects the ranks of the three
corresponding flattenings of $P$. This will lead to explicit
polynomials (i.e., \emph{phylogenetic invariants}) whose vanishing
can be used to identify the tree topology from which  $P$ arises,
for generic parameters.

\begin{thm} \label{thm:4treeid}
Suppose $\lambda < \kappa^2$. With $P$ denoting a
$\kappa\times\kappa\times\kappa\times\kappa$ tensor of
indeterminants, let $S_1$, $S_2$, and $S_3$ denote the sets of
$(\lambda+1)$-minors of the $\kappa^2\times\kappa^2$ matrices
$\Flat_{ab|cd}(P)$, $\Flat_{ac|bd}(P)$, and $\Flat_{ad|bc}(P)$,
respectively. Let the varieties $Y_i=V(S_i)\subset \C^{\kappa^4}$ be
their zero sets. Then

\begin{enumerate}
\item[(i)] $V_{T_i} \subseteq Y_j$ if, and only if, $i=j$.

\item[(ii)] If $P\in V_{T_1}\cup V_{T_2}\cup V_{T_3}$ and $P\in
Y_i\smallsetminus (Y_j\cup Y_k)$ for distinct $i,j,k$, then $P\in
V_{T_i}\smallsetminus (V_{T_j}\cup V_{T_k})$.

\item[(iii)] For distinct $i,j,k$, let $X_i=\phi_{T_i}^{-1}(Y_j\cup
Y_k)\subsetneq \C^L$. Then $X_i$ is a proper algebraic subvariety of
the complex parameter space for the $4$-taxon tree $T_i$, and for
any parameters $s\in\C^L\smallsetminus X_i$ the tree $T_i$ is
identifiable from the joint distribution tensor $P=\phi_{T_i}(s)$
via the vanishing of the polynomials in $S_i$.

\end{enumerate}

\end{thm}

\begin{proof} Throughout, we may assume $i=1,j=2,k=3$.

To establish (i), for any parameters $s\in \C^L$ on $T_1$, let
$Q=Q(s)$, $P=P(s)=\phi_{T_1}(s)$, where $Q$ and $P$ are the tensors
described above. Now $\Flat_{ab|cd}(Q)$ is a matrix of all zeros
except for a single $\lambda\times\lambda$ submatrix whose entries
are those of $A=\diag(\boldsymbol \pi_r)M_{rf}$. Thus
$$\rank(\Flat_{ab|cd}(Q))=\rank(A)\le \lambda.$$ By Equation
(\ref{eq:FP1}), this implies $\rank(\Flat_{ab|cd}(P))\le \lambda$.
Thus $\phi_{T_1}(\C^L)\subseteq Y_1$, and hence $V_{T_1}\subseteq
Y_1$.

We now show $V_{T_1}\not\subseteq Y_2$ or $Y_3$, by finding an
$s\in\C^L$ with $\phi_{T_1}(s)\notin Y_2\cup Y_3$. To pick such an
$s$, we choose each of $M_{ra}, M_{rb}, M_{fc}, M_{fd}$ to have the
block form $( I_{\kappa\times\kappa}\  0_{(\lambda-\kappa)\times
\kappa})^T$. Then the Kronecker products
$N_{ac},N_{bd},N_{ad},N_{bc}$ all have block form
$$(I_{\kappa^2\times\kappa^2}\ 0_{(\lambda^2-\kappa^2)\times
\kappa^2})^T.$$ Now choosing $\boldsymbol \pi_r$ and $M_{rf}$ to
have all positive entries, for instance, ensures that all entries of
$A$ are non-zero. Since $\Flat_{ac|bd}(Q)$ is a matrix of all zeros,
except the entries of $A$ which appear in the $((i,j),(i,j))$
positions,  $\Flat_{ac|bd}(Q)$ is thus a diagonal matrix of rank
$\lambda^2$.  Similarly $\Flat_{ad|bc}(Q)$ is diagonal with full
rank $\lambda^2$. Thus
\begin{align*}
\Flat_{ac|bd}(P)=N_{ac}^T \Flat_{ac|bd}(Q)N_{bd},\\
\Flat_{ad|bc}(P)=N_{ad}^T \Flat_{ad|bc}(Q)N_{bc},\end{align*} both
have rank $\kappa^2>\lambda$ due to the particular form of
$N_{ac},N_{bd},N_{ad},N_{bc}$. Thus $\phi_{T_1}(s)\notin Y_2$ or
$Y_3$.

Statement (ii) follows immediately from (i).

For (iii), note that the existence of the point $s$, constructed
above, with $\phi_{T_1}(s)\notin Y_2$ or $Y_3$ shows $X_1$ is a
proper subset of $\C^L$. That it is an algebraic variety follows
from its definition as the zero set of all polynomials of the form
$f\circ \phi_{T_1}$ where $f$ vanishes on $Y_2\cup Y_3$.
\end{proof}

\begin{rem} The set $X_i$ should be thought of as the set of `bad'
parameters for the model on $T_i$, for which this approach is unable
to identify the tree topology from the resulting joint distribution.
It is important that $X_i$ be a proper subvariety of the parameter
space since this immediately implies its dimension is smaller than
that of the parameter space. If one restricts attention from complex
to real parameters, or even to stochastic parameters, the points in
$X_i$ still form a set of lower dimension than the full parameter
space. Thus for `most' stochastic parameters, the topology is
identifiable from the joint distribution.
\end{rem}

\begin{cor} \label{cor:treeid}
The $n$-taxon bifurcating tree topology is identifiable for generic
parameters of the $(\lambda,\kappa)$-state general Markov model when
$\lambda <\kappa^2$. That is for each $n$-leaf tree $T$, there
exists a proper subvariety $X_T$ of the complex parameter space
$\C^L$ such that the tree topology is identifiable from the joint
distribution arising from any parameter choice $s\in
\C^L\smallsetminus X_T$ via the vanishing of certain explicit
polynomials (to be described below).
\end{cor}

\begin{proof} As is well known \cite{MR2060009}, to identify the topology
of an $n$-leaf tree $T$, it is enough to identify the topology of
each induced quartet tree relating four leaves of $T$.

Let $\C^L$ be the complex parameter space for the tree $T$, and let
$\mathcal Q$ denote the collection of all 4-leaf trees induced by
$T$. For each $T'\in \mathcal Q$, the parameter space for $T'$ is
$\C^{L'}$ and we have the following commutative diagram of
polynomial maps:
$$ \begin{CD}
\C^L @>{\phi_T}>>\C^{\kappa^n}\\
 @V\alpha_{T'} VV @V\mu_{T'} VV \\
\C^{L'} @>{\phi_{T'}}>> \C^{\kappa^4}\end{CD}. $$ The map
$\alpha_{T'}$ can be explicitly given by multiplication of matrix
parameters for $T$ to obtain matrix parameters for $T'$, once a
consistent choice of roots for $T'$ and $T$ is made. The map
$\mu_{T'}$ is a marginalization map on tensors, where we sum over
all but the 4 indices corresponding to leaves of $T'$.

For any $T'\in \mathcal Q$, identify its leaves with labels
$a,b,c,d$ so that $T'$ is identified with $T_1$ of Theorem
\ref{thm:4treeid}. Letting $X_i,Y_i$ be the varieties defined in
that theorem, consider varieties
\begin{align*}
X_T&=\bigcup_{T'\in \mathcal Q} \alpha_{T'}^{-1} (X_1),\\
Y_T&=\bigcap_{T'\in \mathcal Q} \mu_{T'}^{-1}(Y_1).
\end{align*}
For any parameters $s\in \C^L\smallsetminus X_T$, $\phi_T(s)\in
Y_T\smallsetminus \phi_T(X_T)$, and all 4-leaf induced tree
topologies are identifiable by the vanishing of the polynomials
defining $Y_T$. (An explicit set of polynomials defining $Y_T$ can
be taken to be the composition of the polynomials in $S_1$ of
Theorem 2 with the marginalization maps $\mu_{T'}$.)

It only remains to show that $X_T$ is a proper subvariety of $\C^L$.
But it is easy to see that each of the maps $\alpha_{T'}$ is
surjective. Therefore since $X_1\subsetneq\C^{L'}$, for each $T'\in
\mathcal Q$ we find $\alpha_{T'}^{-1}(X_1)\subsetneq\C^L$. Since
$X_T$ is a finite union of proper subvarieties of $\C^L$, we obtain
$X_T\subsetneq \C^L$.
\end{proof}

We note that the idea of using rank conditions on flattenings of a
data tensor to identify tree topology also appears in recent
independent work of Eriksson \cite{Erik}, where the SVD is used to
give a novel algorithm for tree construction. That paper deals only
with the general Markov model ($\lambda=\kappa$), and takes a
slightly different approach to identifying splits for a tree without
focusing on quartets.
\medskip

Specializing our result to $\lambda=\kappa$, we recover the
following result proved previously by Steel \cite{S94} using the
log-det distance, and then reproved by Eriksson.

\begin{cor}
The tree topology is identifiable for generic parameters in the
$\kappa$-state general Markov model.
\end{cor}

\section{Identifiability of tree topology for
analytic $(\lambda,\kappa)$-state models}\label{sec:analyticident}

To deduce identifiability of the tree topology for analytic
$(\lambda,\kappa)$-state submodels using Theorem \ref{thm:4treeid}
or Corollary \ref{cor:treeid} requires a little additional work,
since, \emph{a priori}, it is possible that the restricted
parameters are not sufficiently generic to preserve identifiability.

\medskip

Fix an analytic $(\lambda,\kappa)$-state model, and let $\psi:U\to
\C^L$ denote its Markov map, giving parameters for the general
$(\lambda,\kappa)$-state model in terms of parameters for the
analytic model.

We first consider the identifiability of a 4-leaf tree topology, so
let $T_1=T_{ab|cd}$. For parameters $u$ of the analytic  model, we
need only show that points $\psi(u)$ are generically not in the
variety $X_1$ of Theorem \ref{thm:4treeid}.

With $I(X_1)$ denoting the ideal of all polynomials vanishing on
$X_1$, consider the set
$$\widetilde X_1 =\psi^{-1}(X_1)=\{u : f\circ\psi(u)=0,\text{ for all }f\in I(X_1)\}
\subseteq U\subseteq\R^M.
$$
Since $\psi$ is an analytic map, so is $f\circ\psi$ for each
polynomial $f$, and so $\widetilde X_1$ is an analytic subvariety of
$U$. Thus if we establish that $\widetilde X_1$ is a proper
subvariety of $U$, then generic points in $U$ are mapped to generic
points (off of $X_1$) in $\C^L$ by $\psi$. This establishes the
following:

\begin{lem}\label{thm:anid}
For $\lambda<\kappa^2$, consider an analytic
$(\lambda,\kappa)$-state model on a $4$-leaf tree, with parameter
space $U$ and Markov map $\psi$. If there is a single choice of
parameters $u\in U$ such that $\psi(u)\notin X_1$, then the $4$-leaf
tree topology is identifiable for generic parameters.
\end{lem}

For ease of application to the specific models listed in Section
\ref{sec:submodels}, we deduce a weaker form of this.

\begin{lem} \label{thm:baseclassid} Consider an analytic $\kappa$-base, $m$-class
model on a $4$-leaf tree. Suppose $m<\kappa$ and there is at least
one choice of allowable parameters for which
\begin{enumerate}
\item[(i)] the Markov matrices for pendant edges are of the form
$M_e=M_0=(I_{\kappa\times\kappa}\thickspace I_{\kappa\times\kappa}
\dots\, I_{\kappa\times\kappa})^T,$ and \item[(ii)] if $\boldsymbol
\pi_r$ is the root distribution and $M_e$ the $\lambda\times\lambda$
Markov matrix assigned to the internal edge of the tree, then the
$\kappa\times\kappa$ matrix
$$N=M_0^T\diag(\boldsymbol \pi_r)M_e M_0$$
has at least $m\kappa+1$ non-zero entries. \end{enumerate}
Then the
tree topology is identifiable for generic parameters of the model.
\end{lem}

Before proving this, we note that condition (i) means that no base
changes occur on pendant edges, though class information is hidden.
In condition (ii), $N$ represents a joint distribution of bases,
without class information, at the two internal nodes of the tree.

\begin{proof} Since $m<\kappa$, then $m\kappa<\kappa^2$ and Lemma
\ref{thm:anid} applies.

For a parameter choice on the tree $T_1$ as described in the
statement of the lemma, the joint distribution of bases at the
leaves is given by $P$ where
$$P(i,j,k,l)=\begin{cases}
N(i,k) &\text{ if $i=j$, $k=l$}\\
0 &\text{ otherwise}
\end{cases}.$$

Therefore the matrices $\Flat_{ac|bd}(P)$ and $\Flat_{ad|bc}(P)$ are
diagonal with at least $m\kappa+1$ non-zero entries. Hence they have
rank at least $m\kappa+1$. This shows the parameters do not lie in
$X_1$, and so the topology is identifiable for generic parameters.
\end{proof}

\medskip

We now obtain the result that provided our original motivation for
this work.
\begin{cor}
For the covarion model of Tuffley-Steel, if $\kappa\ge 3$ the
$4$-leaf tree topology is identifiable for generic parameters.
\end{cor}

\begin{proof}
This model is an analytic $\kappa$-base, $2$-class model, and so we
need $\kappa\ge 3$ to apply Lemma \ref{thm:baseclassid}.

For any $R=(R_{ij}), \boldsymbol \pi, s_1,s_2$ with all $s_i,
\pi_i,R_{ij}>0$ for $i\ne j$, the matrix $\diag(\boldsymbol
\pi_r)\exp(Qt_e)$ has all positive entries as long as $t_e>0$, so
then the matrix $N$ has all positive entries. Picking such
parameters, with $t_e>0$ for the internal edge of the tree, and
$t_e=0$ for all pendant edges, Lemma \ref{thm:baseclassid} gives the
result.
\end{proof}

This result includes the $\kappa=4$, $20$ cases which apply to DNA
and protein sequences. Note, however, that the identifiability of
the tree topology for the $\kappa=2$ covarion model remains an open
question.

Finally, the result extends to trees with more than 4 leaves, by an
argument analogous to that of Corollary \ref{cor:treeid}.

\begin{cor}
For the covarion model of Tuffley-Steel, if $\kappa\ge 3$ then
bifurcating tree topologies are identifiable for generic parameters.
\end{cor}

Though we omit the details, one similarly sees that tree topologies
for the 4-base, $m$-class covarion model SSRV of \cite{Galt} are
generically identifiable provided $m<4$. Note that the
implementation of the SSRV in inference software described in that
paper actually had $m=4$, a case not covered by our theorem. It
would of course be desirable to prove identifiability for that case,
and larger $m$, as well.

\medskip
Finally, we can apply this approach to non-covarion rate-variation
models with a finite number of rate classes. As an example, we give
the following result.

\begin{cor}
For the GM+GM+$\cdots$+GM model, with $\kappa$ states and $m$
classes where $m<\kappa$, bifurcating tree topologies are
identifiable for generic parameters. In particular, when $\kappa=4$,
the tree topology is generically identifiable for the GM+GM+GM
model.
\end{cor}
\begin{proof}
For the 4-leaf tree, consider any parameter choice where no
substitutions occur on pendant edges in any of the classes, the root
distribution has all positive entries, all Markov matrix entries are
non-negative, and for at least one class the $\kappa\times\kappa$
Markov matrix for that class on the internal edge has at least
$\kappa+1$ positive entries. Then apply Lemma \ref{thm:baseclassid}.

An argument analogous to that for Corollary \ref{cor:treeid} extends
the result to trees with more leaves.
\end{proof}

Similarly, for the GTR+rate-classes model we obtain generic
identifiability of tree topology provided the number of classes $m$
is less than the number of bases $\kappa$. Note that while previous
result on identifiability for this model \cite{WadSt97, Rog01} have
allowed a known continuous distribution of rates, they have also
assumed a common rate matrix for all classes. Our result holds even
for a model in which different classes have unrelated GTR rate
matrices.

Finally, we note this approach proves generic identifiability of
tree topologies for the GM+I model when $\kappa\ge3$. However, for
this particular model we will take a different approach in another
paper \cite{ARGMI}, obtaining identifiability for $\kappa\ge 2$ as
well as some interesting explicit formulas for recovering
proportions of invariable sites, and identifying other numerical
parameters as well.

\nocite{MEP}

\bibliographystyle{alpha}
\bibliography{Covarion}

\def\cprime{$'$}
\begin{thebibliography}{PMCH01}

\bibitem[AR03]{AR03}
Elizabeth~S. Allman and John~A. Rhodes.
\newblock Phylogenetic invariants for the general {M}arkov model of sequence
  mutation.
\newblock {\em Math. Biosci.}, 186:113--144, 2003.

\bibitem[AR04]{ARQuart}
Elizabeth~S. Allman and John~A. Rhodes.
\newblock Quartets and parameter recovery for the general {M}arkov model of
  sequence mutation.
\newblock {\em App. Math. Res. Express (AMRX)}, 2004:4:107--131, 2004.

\bibitem[AR05a]{ARGMI}
Elizabeth~S. Allman and John~A. Rhodes.
\newblock {Identifying topologies and parameters for the GM+I model}.
\newblock 2005.
\newblock in preparation.

\bibitem[AR05b]{ARgm}
Elizabeth~S. Allman and John~A. Rhodes.
\newblock Phylogenetic ideals and varieties for the general {M}arkov model,
  2005.
\newblock in preparation, {\tt arXiv:math.AG/0410604}.

\bibitem[AR05c]{ARSBD}
Elizabeth~S. Allman and John~A. Rhodes.
\newblock {Phylogenetic invariants for stationary base composition}.
\newblock {\em J. Symbolic Comp.}, 2005.
\newblock to appear, {\tt arXiv:q-bio.PE/0407035}.

\bibitem[Baa98]{MR1664261}
Ellen Baake.
\newblock What can and what cannot be inferred from pairwise sequence
  comparisons?
\newblock {\em Math. Biosci.}, 154(1):1--21, 1998.

\bibitem[BGP05]{MEPLikelihood}
David Bryant, Nicolas Galtier, and Marie-Anne Poursat.
\newblock Likelihood caclulation in molecular phylogenetics.
\newblock In Olivier Gascuel, editor, {\em Mathematics of Evolution and
  Phylogeny}, pages 33--62. Oxford University Press, 2005.

\bibitem[Bun71]{Bun}
Peter Buneman.
\newblock The recovery of trees from measures of dissimilarity.
\newblock In {\em Mathematics in the Archeological and Historical Sciences},
  pages 387--395, Edinburgh, 1971. Edinburgh University Press.

\bibitem[CF87]{CF87}
James~A. Cavender and Joseph Felsenstein.
\newblock Invariants of phylogenies in a simple case with discrete states.
\newblock {\em J. of Class.}, 4:57--71, 1987.

\bibitem[CGS05]{CasGarSul}
Marta Casanellas, Luis~David Garcia, and Seth Sullivant.
\newblock Catalog of small trees.
\newblock In Lior Pachter and Bernd Sturmfels, editors, {\em Algebraic
  Statistics for Computational Biology}, pages 291--304. Cambridge University
  Press, 2005.

\bibitem[CH91]{CH91}
J.~T. Chang and J.~A. Hartigan.
\newblock Reconstruction of evolutionary trees from pairwise disbributions on
  current species.
\newblock In E.~M. Keramidas, editor, {\em Computing Science and Statistics:
  Proceedings of the 23rd Symposium on the Interface}, pages 254--257.
  Interface Foundation, 1991.

\bibitem[Cha96]{MR97k:92011}
Joseph~T. Chang.
\newblock Full reconstruction of {M}arkov models on evolutionary trees:
  identifiability and consistency.
\newblock {\em Math. Biosci.}, 137(1):51--73, 1996.

\bibitem[CHHP00]{CHHP}
B.~Chor, M.~D. Hendy, B.~R. Holland, and D.~Penny.
\newblock Multiple maxima of likelihood in phylogenetic trees: an analytic
  approach.
\newblock {\em Mol. Bio. and Evol.}, 17:1529--1541, 2000.

\bibitem[CKS03]{CKS}
B.~Chor, A.~Khetan, and S.~Snir.
\newblock Maximum likelihood on four taxa phylogenetic trees: Analytic
  solutions.
\newblock {\em RECOMB'03}, 2003.

\bibitem[CS05]{CasSul}
Marta Casanellas and Seth Sullivant.
\newblock The strand symmetric model.
\newblock In Lior Pachter and Bernd Sturmfels, editors, {\em Algebraic
  Statistics for Computational Biology}, pages 305--321. Cambridge University
  Press, 2005.

\bibitem[Eri05]{Erik}
Nicholas Eriksson.
\newblock Tree construction using singular value decomposition.
\newblock In Lior Pachter and Bernd Sturmfels, editors, {\em Algebraic
  Statistics for Computational Biology}, pages 347--358. Cambridge University
  Press, 2005.

\bibitem[ERSS04]{math.AG/0407033}
Nicholas Eriksson, Kristian Ranestad, Bernd Sturmfels, and Seth Sullivant.
\newblock Phylogenetic algebraic geometry.
\newblock 2004.
\newblock to appear, in proceedings of ``Projective Varieties with Unexpected
  Properties,'' Siena, Italy,{ \tt arXiv:math.AG/0407033}.

\bibitem[ES93]{MR93m:62121}
Steven~N. Evans and T.~P. Speed.
\newblock Invariants of some probability models used in phylogenetic inference.
\newblock {\em Ann. Statist.}, 21(1):355--377, 1993.

\bibitem[EW04]{EW04}
Steven~N. Evans and Tandy Warnow.
\newblock Unidentifiable divergence times in rates-across-sites models.
\newblock 2004.
\newblock {\tt arXiv:q-bio.PE/0408011}.

\bibitem[FM70]{FM}
Walter~M. Fitch and Etan Markowitz.
\newblock An improved method for determining codon variability in a gene and
  its application to the rate of fixation of mutations in evolution.
\newblock {\em Biochemical Genetics}, 4:579--593, 1970.

\bibitem[Gal01]{Galt}
Nicolas Galtier.
\newblock Maximum-likelihood phylogenetic analysis under a covarion-like model.
\newblock {\em Mol. Biol. Evol.}, 18(5):866--873, 2001.

\bibitem[Gas05]{MEP}
Olivier Gascuel, editor.
\newblock {\em Mathematics of Evolution and Phylogeny}.
\newblock Oxford University Press, Oxford, 2005.

\bibitem[Hen89]{Hen89}
Michael~D. Hendy.
\newblock The relationship between simple evolutionary tree models and
  observable sequence data.
\newblock {\em Systematic Zoology}, 38:310--321, 1989.

\bibitem[Lak87]{Lake87}
J.A. Lake.
\newblock A rate independent technique for analysis of nucleic acid sequences:
  Evolutionary parsimony.
\newblock {\em Mol. Bio. Evol.}, 4:167--191, 1987.

\bibitem[PMCH01]{PMCH}
David Penny, Bennet~J. McComish, Michael~A. Charleston, and Michael~D. Hendy.
\newblock Mathematical elegance with biochemical realism: The covarion model of
  molecular evolution.
\newblock {\em J. Mol. Evol.}, 53:711--723, 2001.

\bibitem[PS05]{ASCB}
Lior Pachter and Bernd Sturmfels, editors.
\newblock {\em Algebraic Statistics for Computational Biology}.
\newblock Cambridge University Press, Cambridge, 2005.

\bibitem[Rog01]{Rog01}
James~S. Rogers.
\newblock Maximum likelihood estimation of phylogenetic trees is consistent
  when substitution rates vary according to the invariable sites plus gamma
  distribution.
\newblock {\em Syst. Biol.}, 50(5):713--722, 2001.

\bibitem[SHP98]{MR2000e:92016}
Mike Steel, Michael~D. Hendy, and David Penny.
\newblock Reconstructing phylogenies from nucleotide pattern probabilities: a
  survey and some new results.
\newblock {\em Discrete Appl. Math.}, 88(1-3):367--396, 1998.

\bibitem[SS03]{MR2060009}
Charles Semple and Mike Steel.
\newblock {\em Phylogenetics}, volume~24 of {\em Oxford Lecture Series in
  Mathematics and its Applications}.
\newblock Oxford University Press, Oxford, 2003.

\bibitem[SS05]{q-bio.PE/0402015}
Bernd Sturmfels and Seth Sullivant.
\newblock {Toric ideals of phylogenetic invariants}.
\newblock {\em J. Comput. Biol.}, 12(2):204--228, 2005.
\newblock {\tt arXiv:q-bio.PE/0402015}.

\bibitem[SSE93]{MR1218244}
L.~A. Sz{\'e}kely, M.~A. Steel, and P.~L. Erd{\H{o}}s.
\newblock Fourier calculus on evolutionary trees.
\newblock {\em Adv. in Appl. Math.}, 14(2):200--210, 1993.

\bibitem[SSH94]{SSH94}
M.A. Steel, L.~Sz\'ekely, and M.D. Hendy.
\newblock Reconstructing trees from sequences whose sites evolve at variable
  rates.
\newblock {\em J. Comput. Biol.}, 1(2):153--163, 1994.

\bibitem[Ste94]{S94}
M.~Steel.
\newblock Recovering a tree from the leaf colourations it generates under a
  {M}arkov model.
\newblock {\em Appl. Math. Letters}, 7(2):19--23, 1994.

\bibitem[TS98]{MR1604518}
Chris Tuffley and Mike Steel.
\newblock Modeling the covarion hypothesis of nucleotide substitution.
\newblock {\em Math. Biosci.}, 147(1):63--91, 1998.

\bibitem[WS97]{WadSt97}
Peter~J. Waddell and M.A. Steel.
\newblock General time-reversible distances with unequal rates across sites:
  Mixing ${\Gamma}$ and inverse {G}aussian distributions with invariant sites.
\newblock {\em Mol. Phylo. Evol.}, 8(3):398--414, 1997.

\end{thebibliography}

\end{document}